\documentclass[12pt]{article}

\usepackage{epsfig,multicol,multirow,cite}
\usepackage{amsmath,subfigure,latexsym,amssymb}
\usepackage[figuresright]{rotating}
\usepackage{tikz, graphicx}

\usepackage{color,xcolor}
\definecolor{deepfuchsia}{rgb}{0.76, 0.33, 0.76}
\definecolor{pinegreen}{HTML}{008B72}

\usepackage[colorlinks=true,linkcolor=pinegreen,citecolor=magenta,urlcolor=blue]{hyperref}
\usepackage{hyperref}

\newcommand{\be}{\begin{equation}}
\newcommand{\ee}{\end{equation}}
\newcommand{\nn}{\nonumber}
\newcommand{\bea}{\begin{eqnarray}}
\newcommand{\eea}{\end{eqnarray}} 
\newcommand{\eps}{\epsilon}
\newcommand{\la}{\langle}
\newcommand{\ra}{\rangle}

\newcommand{\Z}{\mathbb{Z}}
\newcommand{\R}{{\kern+.25em\sf{R}\kern-.78em\sf{I} \kern+.78em\kern-.25em}}
\newcommand{\RR}{{\kern+.25em\sf{R}\kern-.6em\sf{I} \kern+.6em\kern-.25em}}
\newcommand{\N}{{\kern+.25em\sf{N}\kern-.78em\sf{I} \kern+.78em\kern-.25em}}
\newcommand{\C}{\mathbb{C}}

\newcommand{\ri}{{\rm i}}

\makeatletter
\@addtoreset{equation}{section}
\makeatother

\begin{document}
 
\begin{center}
{\Large\bf Structure of cosmic strings for a gauged}

\vspace*{6mm}

{\Large\bf $B-L$ symmetry and two Higgs fields} \\

\vspace*{1cm}

Jos\'{e} Antonio Garc\'{\i}a-Hern\'{a}ndez,
Victor Mu\~{n}oz-Vitelly \vspace*{0.7mm} \\ and Wolfgang Bietenholz
\\
\ \\
Instituto de Ciencias Nucleares \vspace*{0.7mm} \\
Universidad Nacional Aut\'{o}noma de M\'{e}xico \vspace*{0.7mm} \\
A.P.\ 70-543, C.P.\ 04510 Ciudad de M\'{e}xico, Mexico\\

\end{center}

\vspace*{6mm}

\noindent
In the Standard Model, the difference between the baryon number $B$
and the lepton number $L$ is conserved, which represents a symmetry
that is global and exact, and therefore unnatural.
We turn it into a naturally exact, local symmetry by coupling the
quantum number $B-L$ to an Abelian gauge field.
Gauge anomalies are cancelled by adding
right-handed neutrinos $\nu_R$, which are not sterile in this case.
Therefore the usual Majorana term is forbidden by gauge symmetry,
but we arrange for a $\nu_R$-mass by adding a Higgs-type 1-component
complex scalar field. Thus we arrive at a modest, well-motivated
extension of the Standard Model.

In this framework, we investigate the field equations in the extended
gauge-Higgs sector, involving both Higgs fields and the non-standard
U(1) gauge field. They are given by a set of coupled, non-linear
differential equations, which we solve numerically, focusing on
the structures of cosmic strings. For a variety of parameters, we
identify the string profiles and energy densities, assuming
appropriate boundary conditions.
In particular, these profiles depend on the winding number of each
of the Higgs fields. As an amazing peculiarity, we discover --- for
high winding numbers --- ``overshooting'' and ``co-axial'' solutions;
in the latter case, the profile of the standard Higgs field changes
its sign near the core of the cosmic string.

\newpage

\tableofcontents

\section{Motivation}
\label{moti}
The Standard Model of particle physics is extremely successful:
its predictions have been confirmed over and over again, to
higher and higher precision. Yet, there are reasons not to be
completely satisfied with it: of course, it does not include
gravity, Dark Matter and Dark Energy. However, in this work
we take an intrinsic, conceptual shortcoming as our point of
departure: the exact conservation of the difference between the
baryon number $B$ and lepton number $L$. The Standard Model
allows for individual changes of $B$ and of $L$, due to topological
windings in the SU(2)$_L$ gauge field. However, the theoretical
transition rate is extremely low (at accessible energies), hence
$B$ and $L$ number violations have never been observed.

On the other hand, the conservation of the difference $B-L$ is
theoretically exact, as reflected by the property that the baryon
current $J_{\mu}^{\rm B}$ and the lepton current $J_{\mu}^{\rm L}$
have the same divergence,
\be  \label{divergence}
\partial^{\mu} J_{\mu}^{\rm B} =
\partial^{\mu} J_{\mu}^{\rm L} =
- \frac{N_{\rm g}}{16 \pi^{2}} {\rm Tr} [ W^{\mu \nu} \tilde W_{\mu \nu} ] \ , 
\ee
where $N_{\rm g}$ is the number of fermion generations, while
$W^{\mu \nu}$ and $\tilde W_{\mu \nu} = \tfrac{1}{2} \epsilon_{\mu \nu \rho \sigma}
W^{\rho \sigma}$ are the ${\rm SU}(2)_L$ field strength tensor and
its dual, see {\it e.g.}\ Ref.\ \cite{PesSch}.

Hence $B-L$ conservation represents an exact, global symmetry of
the Standard Model, $\partial^{\mu} (J_{\mu}^{\rm B} - J_{\mu}^{\rm L})=0$.
This seems unnatural: global symmetries are usually just approximately
valid in some energy regime --- in general, there is no reason
for them to be exact. Another exception within the Standard
Model is Lorentz invariance, which --- along with locality --- also
implies CPT symmetry \cite{CPT}. However, from the perspective of
General Relativity, Lorentz invariance can be interpreted as a gauge
symmetry, so from that point of view its exactness is natural
and necessary.

In this work, we refer to the analogous scenario for the $B-L$
invariance, which we turn into a (local) gauge symmetry. So we
step beyond the Standard Model, but
in a controlled and economic manner, by adding only few fields,
for which there is a crystal-clear motivation.

First, this means that we couple the charge $B-L$ to an Abelian
gauge field ${\cal A}_{\mu}$, which explains its conservation.
The Higgs mechanism will arrange for this gauge field to be heavy,
thus avoiding long-range effects and therefore also avoiding
obvious contradictions with observations. This addition alone leads
to gauge anomalies, due to the unbalance between the left- and
right-handed fermion content of the Standard Model. This is fixed
by adding a right-handed neutrino $\nu_R$ to each fermion
generation (our notation does not distinguish its flavor).

Usually, this addition provides a Majorana mass term, which is,
however, forbidden in this scenario, because here such a term has
a gauge charge $B-L=-2$. Conceptually, it is not strictly required to
obtain a mass for the right-handed neutrino. Still, we consider it well
motivated to enable such a mass, by means of an extra Higgs-type
field $\chi$. Its form is again minimal, just a 1-component, complex
scalar field. These are all the ingredients that we add to the
Standard Model.

Once the right-handed neutrino $\nu_R$ has mass, we assume it to be
very heavy, say above $10^{20}\,$eV, such that (a variant of) the seesaw
mechanism applies, as an explanation why $\nu_L$ is so light.
This also justifies the fact that $\nu_R$ has not been observed,
although in this scenario it couples to the gauge field ${\cal A}_{\mu}$,
in addition to the Yukawa-type coupling to the standard and non-standard
Higgs field.

Within this gauge-Higgs sector, we study the field equations, which
can be solved numerically. We are interested in vortices
in the ${\cal A}_{\mu}$ configurations as topological defects,
and the resulting non-standard cosmic strings.\footnote{We mean
``non-standard'' just in the sense that fields beyond the Standard
Model are involved.} For a variety of sectors
in parameter space, we explore the radial profile functions of such
local cosmic strings, and their dependence on the winding numbers of
both Higgs fields (standard and non-standard). We present a set of
solutions, which are not ruled out by observations; this statement
includes recent bounds based on gravitational waves \cite{LIGO1,LIGO3}.
Hence such cosmic strings could have persisted since the early Universe,
along the lines suggested in Refs.\ \cite{Kibble}, and reviewed
{\it e.g.}\ in Refs.\ \cite{HindKibVil} (though these references mostly
refer to standard cosmic strings\footnote{Hindmarsh and Kibble also
discuss axionic cosmic strings.}).

As an interesting peculiarity, we identify an extreme region in parameter
space where solutions emerge that we denote as ``overshooting'' and
``co-axial'' strings. ``Overshooting'' means that a profile function
starts a 0 in the core and exceeds at short distances its asymptotic
value far from it. In the co-axial case,
the profile function of the standard Higgs field $\Phi$
changes its sign near the core of the string. Such exotic solutions are
hardly known, although this possibility was reported before in
Ref.\ \cite{Bogo}, but in a different set-up, without $(B-L)$-gauging.

Section \ref{string} reviews some basic aspects of cosmic strings.
In Section \ref{model}, we explicitly describe the (modestly) extended
Standard Model that this work deals with. In particular,
Subsection \ref{fieldeq} discusses the corresponding field equations
in the sector composed of the non-standard Abelian gauge field
${\cal A}_{\mu}$, as well as the standard and non-standard Higgs field,
$\Phi$ and $\chi$. A multitude of numerical solutions, for different
action parameters and winding numbers, is
presented in Section \ref{profile}, before arriving at our summary
and conclusions in Section \ref{conclu}.

Appendix \ref{numeri} adds remarks about the numerical method
that we applied to obtain the results in Section \ref{profile},
including comments on their reliability
and precision. Appendix \ref{bound} discusses the conditions for
the Higgs-Higgs potential to be bounded from below.
Preliminary results of this work were reported before in two
theses \cite{Victor,JA} and a proceeding contribution \cite{proc}.

\section{Cosmic strings}

\label{string}
Topology plays a relevant role in Quantum Field Theory, for
instance the term on the right-hand side of eq.\ (\ref{divergence})
represents the topological density (the Chern-Pontryagin density)
of a configuration of the
gauge field $W_{\mu}$ of the weak interaction. Generally, the
differentiable configurations of an SU($N$) gauge field ($N \geq 2$)
fall into disjoint topological sectors (in four Euclidean dimensions,
with suitable boundary conditions), each one characterized
by a topological charge $Q \in \Z$.

Even when a theory does not have topological sectors from a global
perspective, it may still have local topological defects. They
are well-known in condensed matter systems, such as type II
superconductors \cite{Abrikosov} (where they lead to Abrikosov strings),
non-linear optical systems \cite{Ducci} and superfluid helium
$^{4}$He \cite{Feyn,Lin}.
Some systems undergo phase transitions, which are related to the
percolation of topological defects. This happens in particular
in the 2d XY model, where the (un)binding of vortex--anti-vortex
pairs drives the Berezinskii-Kosterlitz-Thouless phase transition
(for a review, see Ref.\ \cite{WBUrs}).
In higher dimensions, {\it e.g.}\ in the 3d XY model, such topological
defects in 2d sheets tend to pile up to form closed, global strings.

\subsection{Prototypes}

We begin with prototype solutions for global and local cosmic
strings, which are inspired by Refs.\ \cite{HindKibVil}.
The latter will be extended to our system of interest in Section
\ref{profile}.

\subsubsection{Global cosmic strings}
\label{globalcs}

\begin{figure}[h!]
\centering
\vspace*{-2mm}
\includegraphics[scale=0.75]{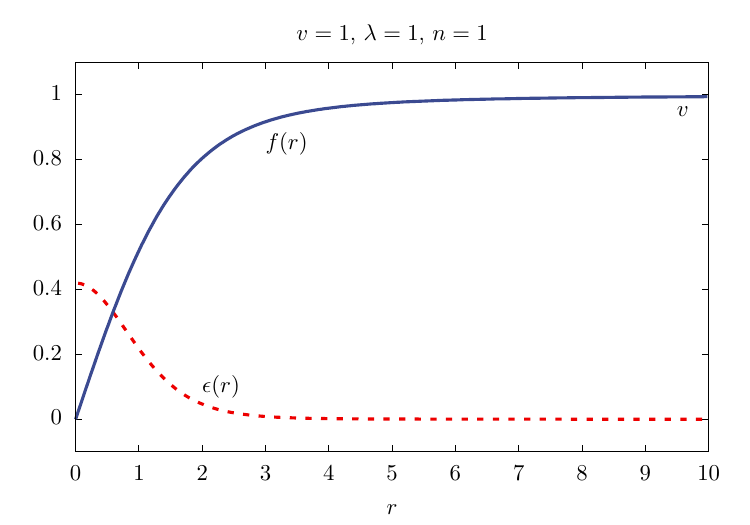}
\vspace*{-3mm}
\caption{\small A prototype profile function $f(r)$ of a global cosmic
string, with a topological defect at $r=0$, and $f(r \to \infty) = v$,
  in cylindrical coordinates, which is obtained by numerically solving
  eq.\ (\ref{protodiff}).
  Here we set $\lambda_0 =1$ and introduce dimensionless units by further
  setting $v=1$, which implies $\mu_0^2 = -v^2 \lambda_0 = -1$.
  This example refers to winding number $n = 1$.
  The energy density $\epsilon (r)$ is computed according to eq.\
  (\ref{protoeps}) up to an additive constant;
  it decays rapidly as $r$ increases.}
\label{protprof}
\vspace*{-5mm}
\end{figure}

Let us illustrate the concept with the simple case of a complex scalar
field $\chi (x) \in \C$ with the Lagrangian
\be  \label{lp4}
{\cal L}(\chi, \partial_{\mu} \chi) = \frac{1}{2} \partial^{\mu} \chi^{*}
\partial_{\mu} \chi - V(|\chi|^{2}) \ , \quad
V(|\chi|^{2}) = \frac{1}{2} \mu_{0}^{2} |\chi|^{2} +
\frac{1}{4} \lambda_{0} |\chi|^{4} \ ,
\ee
with $\lambda_0 > 0$.
For $\mu_{0}^{2} < 0$, there is a continuous set of classical vacua,
which can be parameterized as
\be
\chi_{0} = v e^{\ri \alpha} \ , \quad v = \sqrt{-\mu_{0}^{2}/\lambda_0} \ ,
\quad \alpha \in (-\pi, \pi] \ .
\ee
An obvious ansatz for its generalization to static solutions with
non-minimal energies, in cylindrical coordinates
($r \geq 0,\ \varphi \in  (-\pi, \pi]$, $z \in \R$), reads
\be  \label{proto1}
\chi(r,\varphi,z) = f(r) e^{\ri n \varphi} \ ,
\ee
where $f(r) \in \R$ is the {\em profile function} and $n \in \Z$ the
{\em winding number.}
For the former we impose the boundary conditions $f(r \to \infty)
= v$, and --- for $n \neq 0$ --- $f(0) = 0$, which avoids a phase
ambiguity. These boundary conditions are applied when we numerically
solve the field equation
\be  \label{protodiff}
f''(r) + \frac{1}{r} f'(r) = \Big( \frac{n^{2}}{r^{2}} + \mu_{0}^{2} +
\lambda_{0} f^{2}(r) \Big) f(r) \ .
\ee
A prototype solution, for $n = 1$, is shown in Figure \ref{protprof}.
This plot also shows the radial energy density $\epsilon (r)$, which
--- in this case --- is given by
\bea
\epsilon (r) &=& \frac{1}{2} | \partial_t \chi |^{2} +
\frac{1}{2} |\partial_r \chi |^{2} + \frac{1}{2r^2}
| \partial_\varphi \chi |^{2} +
\frac{1}{2} |\partial_z \chi |^{2} + V(|\chi|^2) \nn \\
&=& \frac{1}{2} (\partial_r f )^{2} + \frac{1}{2}
\Big( \frac{1}{r} n f \Big)^{2} + \frac{1}{2} \mu_{0}^{2} f^{2} +
\frac{1}{4} \lambda_{0} f^{4} \ ,
\label{protoeps}
\eea
where we used the properties that this solution is static and
constant in $z$. In the limits $r \ll 1$ and $r \gg 1$ the solution
can be expanded analytically in terms of Bessel functions, in agreement
with Figure \ref{protprof} \cite{JA}.

\subsubsection{Local cosmic strings}
\label{locprotosubsec}

Since our study is going to address local cosmic strings, we
extend our prototype accordingly, by coupling the scalar field
$\chi$ to a U(1) gauge field $A_{\mu}$.\footnote{In the literature,
this is known as an ``Abelian Higgs model'', which allows for the
formation of ``Nielsen-Olesen strings''.\label{fnAH}}
Thus the Lagrangian
(\ref{lp4}) turns into
\bea
{\cal L}(\chi, A_{\mu}) &=& \frac{1}{2} (D^{\mu}\chi)^{*} (D_{\mu}\chi)
- V(|\chi|) - \frac{1}{4} F^{\mu \nu} F_{\mu \nu} \ , \nn \\
{\rm where} && D_{\mu} \chi = (\partial_{\mu} + \ri g A_{\mu}) \chi \ , \quad
F_{\mu \nu} = \partial_{\mu} A_{\nu} - \partial_{\nu} A_{\mu} \ ,
\label{proto2Lag}
\eea
where $g$ is the gauge coupling. Now the field equations take the
form of two coupled, non-linear differential equations,
\bea
\Big( D^{\mu} D_{\mu} + \mu_{0}^{2} \chi + \lambda_0 |\chi|^{2}\Big) \chi
&=& 0 \ , \nn \\
D^{\mu} F_{\mu \nu} + \frac{\ri g}{2} \Big( (D_{\nu} \chi)^{*} \chi -
\chi^{*} D_{\nu} \chi \Big) &=& 0 \ .
\label{proto2dgl1}
\eea
We stay with the cylindrical ansatz (\ref{proto1}) and add an ansatz for
the configuration of $A_{\mu}$, with $A_z =0$. We fix the gauge such that
$A_{0} = A_{r} =0$, {\it i.e.}\ there is only a component in the direction
of the tangential unit vector $\hat \varphi$,
\be  \label{proto2}
\chi(r,\varphi,z) = f(r) e^{\ri n \varphi} \ , \quad
\vec A (r,\varphi,z) = \frac{a(r)}{r} \hat \varphi \ .
\ee
This ansatz extends the differential equation (\ref{protodiff}) to
\bea  \label{proto2diff}
f'' + \frac{1}{r} f' - \frac{1}{r^{2}} (n + ga)^{2} - \mu_{0}^{2} f
- \lambda_0 f^{3} &=& 0 \ , \nn \\
a'' - \frac{1}{r} a' - g (n + ga) f^{2} &=& 0 \ .
\label{proto2dgl2}
\eea

Ansatz (\ref{proto2}) requires $a(0)=0$, and --- as in
Subsection \ref{globalcs} --- any winding number
$n\neq 0$ also implies $f(0) = 0$.
At asymptotically large $r$, constant solutions are
assumed with $a(r \to \infty) = -n/g$ and (as in the case of the
global string) $f(r \to \infty) = v = \sqrt{- \mu_0^{2} / \lambda_0}$.

\begin{figure}[h!]
\vspace*{-2mm}
\centering
\includegraphics[scale=0.75]{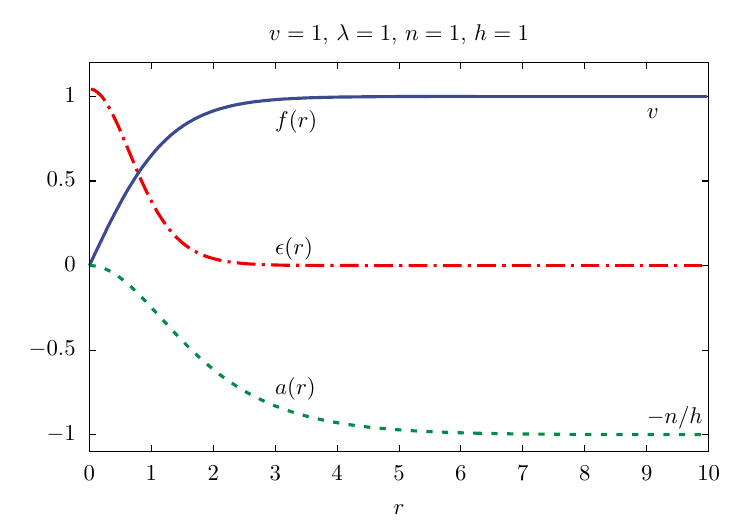}
\vspace*{-3mm}
\caption{\small Profile functions $f(r)$ and $a(r)$ of a prototype local
cosmic string, with a topological defect at $r=0$, in cylindrical
coordinates, obtained by numerically solving eqs.\ (\ref{proto2diff}).
As in Figure \ref{protprof}, we set $\lambda_0 =1$ and $v=1$, which
implies $\mu_0^2 =-1$, and we refer to the winding number $n = 1$.
The energy density $\epsilon (r)$, given in eq.\ (\ref{proto2eps}),
exceeds in the core the value of our global prototype string
due to the gauge field contributions.}
\label{prot2prof}
\vspace*{-2mm}
\end{figure}

Again the cases of small and large $r$ can be analyzed analytically
\cite{JA}, in agreement with the numerical solution shown for a
local prototype in Figure \ref{prot2prof}. In this case, the
energy density is given by
\be  \label{proto2eps}
\epsilon (r) = \frac{1}{2} |\partial_r \chi |^{2} + \frac{1}{2}
\left| \frac{1}{r} \partial_{\varphi} \chi + \frac{\ri g}{r} a \chi
\right|^{2} + \frac{1}{4} F^{\mu \nu} F_{\mu \nu} + V(|\chi|) \ .
\ee

Interestingly, the string's energy per length in the $z$-direction,
$E/z$, which is obtained by integrating over the $(R,\varphi)$-plane
up to a large radius $R$ (with a regularization in the core), diverges
logarithmically for the global string, $E/z \propto \ln R$. In contrast,
$E/z$ remains finite for a local string \cite{HindKibVil}.

\subsection{Formation of cosmic strings}

Cosmic strings could be present throughout the Universe,
according to the mechanism suggested by Kibble \cite{Kibble}.
At the high energies of the Early Universe, topological defects were
naturally omnipresent. Under rapid cooling, part of them might have
persisted, although they would not be expected in a thermalized
low-energy world.

At less than $10^{-12}$~sec after the Big Bang, the temperature
was above 160~GeV, and the electroweak interaction dominated.
This period ended when the Higgs field $\Phi(x) \in \C^{2}$ acquired
a non-zero vacuum expectation value, $| \langle \Phi \rangle | > 0$.
Well separated regions were causally disconnected, hence in such
regions the complex phases of $\Phi(x)$ could not be correlated.
According to Kibble, vortices could have emerged in the interfaces
between such regions. The phase change --- integrated over
a closed loop within such a surface, around a vortex --- amounted
to a non-zero, integer multiple of $2 \pi$. Vortices or anti-vortices
naturally pile up to form strings, which may have led
to a network of cosmic strings, as discussed {\it e.g.}\ in Ref.\
\cite{Brand}. Later they might have lost some energy through
particle emission or gravitational radiation,
but their topological structure could have stabilized
them on long term \cite{James}, to this day.

In the context of superfluid $^{4}$He, where vortex lines behave
like Abrikosov strings \cite{Feyn}, this evolution is known as the
Kibble-Zurek mechanism \cite{Zurek}, which has been experimentally
observed \cite{Lin}. However, the
existence of cosmic strings in the Universe is not confirmed by
observations so far; bounds are set based on the Cosmic Microwave
Background \cite{JS,WPW,CACM} and lately on gravitational wave detections
\cite{LIGO1,LIGO3}. Numerical studies of the Kibble-Zurek mechanism,
by means of Monte Carlo simulations, were conducted in the
XY model in $d=2$ \cite{JelCug} and in $d=3$ \cite{Lat23} spatial
dimensions.

\section{Standard Model extension with a gauged $B-L$ symmetry and
massive, right-handed neutrinos}

\label{model}
Let us present the building blocks of the model under consideration,
in particular its extensions beyond the Standard Model.

As we anticipated in Section \ref{moti}, our point of departure is
promoting $B-L$ invariance to a gauge symmetry, by coupling the
charge $B-L$ to an Abelian gauge group U(1)$_{Y'}$, where $Y'$ takes
a role analogous to the Standard Model weak hypercharge $Y$.
We denote the gauge field that corresponds to U(1)$_{Y'}$ by ${\cal A}_{\mu}$,
cf.\ Section \ref{moti}.
It is coupled to a linear combination of the conserved charges $Y$
and $B-L$, which we write in the form
\be  \label{Yplincomb}
Y' = 2h Y + \frac{1}{2} h' (B-L) \ ,
\ee
where we have introduced the coupling constants $h$ and $h'$ (the
coefficients 2 and 1/2 will turn out to be convenient).

Once this additional gauge field is present, we have to worry about
gauge anomalies. As long as we deal with the gauge charges of the
Standard Model, there are no such anomalies, but now we have the
additional charge $B-L$, which couples to ${\cal A}_{\mu}$, so an
anomaly emerges due to the divergence
of the $(B-L)$-current, $\partial^{\mu} J_{\mu}^{B-L}$.
In fact, for the model that we have assembled so far, this current is
anomalous: in each fermion generation, we have an electrically charged
lepton with both chiralities, plus a left-handed neutrino, which
sum up to $L=3$, with a left-handed dominance.
The quarks, however, contribute two flavors
with both chiralities, 3 colors and $B=1/3$,
which leads to $B=4$, with balanced chiralities.

The overall balance between left and right chiralities
can be attained by adding a right-handed
neutrino $\nu_R$ to each fermion generation. Then we obtain $L=4$ in
each generation, and the divergence of the
$(B-L)$-current vanishes, $\partial^{\mu} J_{\mu}^{B-L} = 0$. 
Since $\nu_R$ is neutral with respect to the gauge charges
of the Standard Model, it does not lead to any gauge anomalies
involving other gauge fields either.\footnote{The scheme of gauging
the $B-L$ symmetry and adding right-handed neutrinos has also been
interpreted as a gauged Peccei-Quinn symmetry \cite{ISY}.}

The fact that models of this kind are free of gauge anomalies, for a
general linear combination (\ref{Yplincomb}) (at least without
involving gravity), is shown in
Refs.\ \cite{Weinberg,ADH03}.\footnote{For the
fermion and scalar charges of the U(1)$_{Y'}$ gauge field, we
refer to Table I in Ref.\ \cite{ADH03}. Without considering gravitons,
the gauge anomaly cancellation conditions leave two degrees of freedom
for these charges, which correspond to our couplings $h$ and $h'$.
If we take into account the anomaly the cancellation conditions which mix
gravitation with U(1)$_{Y'}$ and $[{\rm U}(1)_{Y'}]^{3}$, there are two
more constraints. They are fulfilled as well, if we assume three
fermion generations, and in each generation a neutrino $\nu_{R}$, all
three with the same coupling to U(1)$_{Y'}$ ($-h'/2$ in our notation).}
Now we can build a neutrino mass with a usual Yukawa term, by coupling 
left-handed and right-handed neutrino fields to the standard Higgs field.
Again as usual, this also allows for generation mixing due to the
Pontecorvo-Maki-Nakagawa-Sakata matrix. Thus our model is compatible
with the observed neutrino oscillation.

On the other hand, in contrast
to the usual situation, we cannot add a Majorana mass term for $\nu_R$,
because this term 
$\propto (\bar \nu_R^{\rm T} \ri \sigma_2 \nu_R + {\rm c.c.})$
carries the charge $B-L = -2$. Still, we consider
the model unnatural without a (purely) right-handed neutrino mass
term, impeded by the gauge group U(1)$_{Y'}$.

The simplest way to incorporate such a mass term is the inclusion
of a non-standard Higgs-type field, which is just a 1-component, complex
scalar field $\chi (x) \in \C$, carrying the quantum number $B-L=2$. This
scalar field enables a term of the type
$\, \chi \,  \nu_R^{\rm T} \ri \sigma_2 \nu_R + {\rm c.c.}\,$
in one generation. For several generations another mixing is possible,
in addition to the Pontecorvo-Maki-Nakagawa-Sakata mixing.

The field $\chi$ naturally has a Lagrangian with a quartic potential,
just as in eq.\ (\ref{lp4}), which keeps it power-counting renormalizable.
Moreover, the standard Higgs field $\Phi \in \C^{2}$ and the non-standard
Higgs field $\chi \in \C$ (both with dimension mass) can also
contribute a combined, gauge
invariant, quartic term $\propto \Phi^{\dagger} \Phi \chi^{*} \chi$. \\

So we have added the fields ${\cal A}_{\mu}$, $\nu_R$ and $\chi$ to
the Standard Model, which leads to a consistent, well-motivated
model, and we won't add anything more.
Still, we will occasionally refer to the scenario where this model
is part of a Grand Unified Theory (GUT). Since its gauge group
${\rm SU}(3)_{\rm c} \otimes {\rm SU}(2)_{L} \otimes {\rm U}(1)_{Y}
\otimes {\rm U}(1)_{\rm Y'}$ has rank 5, the GUT gauge group cannot
be SU(5), so the simplest candidate is SO(10) \cite{Mink74}.
In contrast to SU(5), the SO(10) GUT is not ruled out by the lower
bound for the proton lifetime of about $10^{32}$ years.
In this theory, the non-standard hypercharge $Y'$, as a linear
combination of $Y$ and $B-L$, takes the specific form \cite{Buch91}
\be \label{YpSO10}
Y' = Y - \frac{5}{4} (B-L) \ .
\ee
Thus it fulfills the orthogonality condition $\sum_f Y_f Y_f' = 0$,
as well as the normalizations $\sum_f Y_f^2 = 10/3 N_{\rm g}$,
$\sum_f Y_f{'}^2 = 5 N_{\rm g}$, where $N_{\rm g}$ is again the number of
fermion generations, and the sums run over all fermions.
In the numerical analysis of Section \ref{profile}, we are going
to pay attention to the specific form of $Y'$ given in eq.\
(\ref{YpSO10}), with $h'/h = -5$.\\

A study of the field equations of this extended Standard Model,
in its complete form, is beyond the scope of this work. Here we
focus on its gauge-Higgs sector involving the fields $\Phi$,
$\chi$ and ${\cal A}_{\mu}$, which extends the prototype Lagrangian
(\ref{proto2Lag}) to
\be
   {\cal L} = \frac{1}{2} D^{\mu} \Phi^{\dagger} D_{\mu} \Phi +
    \frac{1}{2} d^{\mu} \chi^{*} d_{\mu} \chi - V(\Phi, \chi) - \frac{1}{4}
   {\cal F}_{\mu \nu} {\cal F}^{\mu \nu} \ ,
\ee
where
\bea
   D_{\mu} \Phi &=& \partial_{\mu} \Phi + i h {\cal A}_{\mu} \Phi \ ,
   \quad
   d_{\mu} \chi = \partial_{\mu} \chi + i h' {\cal A}_{\mu} \chi \ , \nn \\
V(\Phi, \chi) &=& \frac{1}{2} \mu^{2} \Phi^{\dagger} \Phi +
\frac{1}{4} \lambda (\Phi^{\dagger} \Phi )^{2} + \frac{1}{2}
\mu'^{\, 2} \chi^{*} \chi + \frac{1}{4}\lambda' (\chi^{*} \chi )^{2} \nn \\
  && + \frac{1}{2} \kappa  \Phi^{\dagger} \Phi \chi^{*} \chi \ .
\label{PotPhiChi}
\eea
Regarding the covariant derivatives $D_{\mu} \Phi$ and $d_{\mu} \chi$,
we considered our ansatz (\ref{Yplincomb}) along with the known
charges $Y_{\Phi} = 1/2$ and $(B-L)_{\chi} = 2$.
On the other hand,
in this analysis we do not include the couplings of $\Phi$ to
the Standard Model gauge fields SU(2)$_L$ and U(1)$_Y$, nor its
Yukawa couplings. Hence we do not consider any of the Standard Model
gauge or fermion fields, assuming that they do not significantly affect
the cosmic string structures to be reported in Section \ref{profile}.

We are interested in the case where both Higgs fields undergo spontaneous
symmetry breaking, at least in the space outside the core of a cosmic
string. In particular, this should lead to a mass of the gauge
field ${\cal A}_{\mu}$, $m_{\cal A} = h v + h'v'$,
which is sufficiently large to explain
that this gauge boson has not been observed.
We denote the corresponding vacuum expectation
values (VEVs) asymptotically far from this core as
\be
v = | \la \Phi (r \to \infty) \ra | \ ,
\quad v' = | \la \chi (r \to \infty) \ra | \ .
\ee
Since our analysis will be kept on tree level, we do not
keep track of renormalization effects, so the conditions $v>0$,
$v'>0$ imply
\be  \label{kappa1constr}
\mu^2 < 0 \ , \quad \mu'^{\, 2} < 0 \ , \quad \kappa <
{\rm min} \Big(\frac{\mu^{2}\lambda'}{\mu'^{\, 2}},
       \frac{\mu'^{\, 2}\lambda} {\mu^{2}} \Big) \ .
\ee
Moreover, for the system to have a ground state, the potential
$V(\Phi, \chi)$ must be bounded from below. This imposes the
additional constraints
\be  \label{kappa1constr2}
\lambda > 0 \ , \quad \lambda' > 0 \ , \quad
\kappa < \sqrt{\lambda \lambda'} \ .
\ee
The constraints on $\kappa$ in eqs.\ (\ref{kappa1constr}) and
(\ref{kappa1constr2}) are not obvious and will be derived in
Appendix \ref{bound}. They can be synthesized by the single inequality
\be  \label{kappa2constr}
\kappa^2 < \lambda \lambda' \ .
\ee

\subsection{Gauge-Higgs field equations}
\label{fieldeq}

In accordance with Section \ref{string}, we search for static solutions
of the field equations, for which we make an ansatz in cylindrical
coordinates, as in eq.\ (\ref{proto2}),
\bea
\Phi (r,\varphi,z) &=& \left( \begin{array}{c} 0 \\ 1 \end{array} \right)
\phi (r) \exp (\ri n \varphi) \ , \nn \\
\chi (r,\varphi,z) &=& \xi (r) \exp (\ri n' \varphi) \ , \nn \\
\vec {\cal A} (r,\varphi,z) &=& \frac{a(r)}{r} \hat \varphi \ .
\label{stringansatz}
\eea
Here $\phi(r)$, $\xi(r)$ and $a(r)$ are the radial profile functions
of a static field configuration, $\hat \varphi$ is the tangential unit
vector, and $n,\, n' \in \Z$ are the winding numbers of the standard
and non-standard Higgs field, respectively.

This ansatz leads to the Euler-Lagrange field equations
\bea
  \frac{d^2 \phi}{dr^2} +\frac{1}{r} \frac{d\phi}{dr} &=&
  \Big[ \frac{(n + h a)^2}{r^2}  + \mu^2 + \lambda \phi^2 +
  \kappa \xi^2 \Big] \phi , \nn \\
  \frac{d^2 \xi}{dr^2} + \frac{1}{r} \frac{d\xi}{dr} &=&
  \Big[ \frac{(n' + h' a)^2}{r^2} + \mu'^{\, 2} + \lambda' \xi^2
   + \kappa \phi^2 \Big] \xi , \nn \\
  \frac{d^2 a}{dr^2}-\frac{1}{r}\frac{da}{dr} &=&
  h \phi^2 (n + ha) + h' \xi^2 (n' + h'a) \ .
\label{fieldeqs}
\eea
Their derivation is non-trivial but straightforward, hence we do
not display intermediate steps here; they are documented in detail
in Refs.\ \cite{Victor,JA}.

Before we proceed to the numerical solutions of eqs.\
(\ref{fieldeqs}), we still have to specify the boundary conditions
for the profile functions. They again generalize the setting of
Subsection \ref{locprotosubsec}:

\begin{itemize}

\item In the core of the string, at $r=0$, we have to require
$a(0)$ to vanish, as the ansatz (\ref{stringansatz}) shows. The same
applies to both Higgs fields, but in that case only if the corresponding
winding number is non-zero, such that a phase ambiguity has to be
avoided. Altogether, the boundary conditions in the core take the form
\be  \label{boundcore}
\phi(0) = 0 \quad {\rm if}~ n\neq 0 \ ; \quad
\xi(0) = 0 \quad {\rm if}~ n'\neq 0 \ ; \quad a(0) = 0 \ .
\ee

\item Asymptotically far from the string, for $r \to \infty$, we require
all three profile functions to be constant in $r$. This implies
that $\phi (r \to \infty)$ and $\xi (r \to \infty)$ converge to
their VEVs in the absence of a string, {\it i.e.}\ $v$ and $v'$,
respectively. Eqs.\ (\ref{fieldeqs}) show that this limit also
relates $v$ and $v'$.

Since these VEVs are independent, $a(r\to \infty)$ has to coincide
with $-n/h$ and with $-n'/h'$, which imposes an additional relation
between the two winding numbers.

To summarize, the asymptotic boundary conditions (in their simplest
form) read
\bea
\phi(r \to \infty) &=& v = \frac{\kappa \mu{'}^{2} - \mu^{2}\lambda'}
{\lambda \lambda' - \kappa^{2}} \ , \nn \\
\xi(r \to \infty) &=& v'= \frac{\mu^{2} \kappa - \mu{'}^{2}\lambda}
{\lambda \lambda' - \kappa^{2}} \ ,
\nn \\
a(r \to \infty) &=& -\frac{n}{h} = - \frac{n'}{h'} \ .
\label{asymcond}
\eea
The inequalities (\ref{kappa1constr}) and (\ref{kappa2constr})
ensure $v, v' > 0$.

\end{itemize}

Usually we will be dealing with dimensionless units. In order to
convert them to physical units, we take $v = 246\, {\rm GeV}$ as
our reference quantity, which turns all other quantities into
physical units as well. In the examples to be reported in Section
\ref{profile}, we consider two values for $v$, which convert the
distance $r=1$ into a physical length as follows:
\bea
v = 0.01 &:& r = 1 \ {\rm corresponds~to} \ 8 \cdot 10^{-6} \, {\rm fm} \nn \\
v = 0.5~  &:& r = 1 \ {\rm corresponds~to} \ 4 \cdot 10^{-4} \, {\rm fm} \ .
\label{units}
\eea

\section{Cosmic string profiles}

\label{profile}
We are now going to present numerical solutions for a variety
of parameter sets. In all these examples, we set $\lambda =
\lambda'=1$; varying these self-couplings did not lead to
particularly instructive insight. Moreover, in all examples
except for the very last one, we set $v' = 1$.
In all the figures, the complete parameter set is given
in the plot title.

\begin{figure}[!h]
\begin{center}
\vspace*{-2mm}
\includegraphics[scale=0.8]{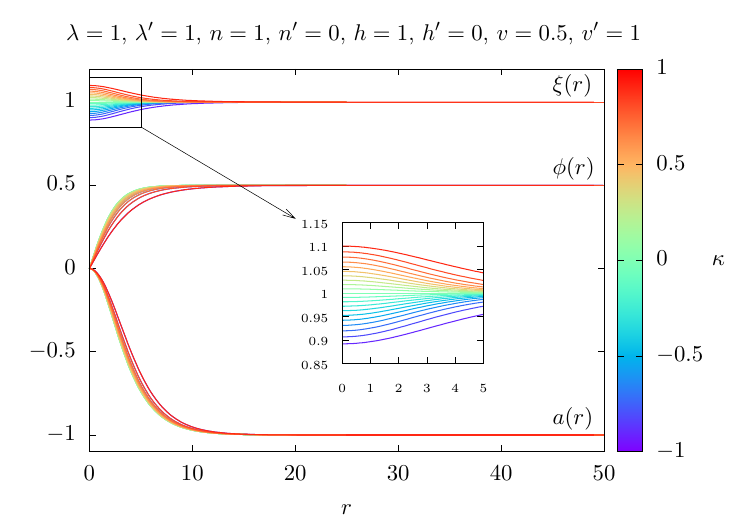}
\vspace*{-3mm}
\caption{\small String profiles for the fields $\phi(r)$, $\xi(r)$
and $a(r)$ over the range $|\kappa | < 1$. Due to $n'=0$, $\xi(0)$
does not vanish,  but there is still the condition $\xi'(0) = 0$.
\label{fign1np0}}
\end{center}
\vspace*{-6mm}
\end{figure}

\begin{figure}[!h]
\vspace*{-2mm}
\begin{center}
\includegraphics[scale=0.8]{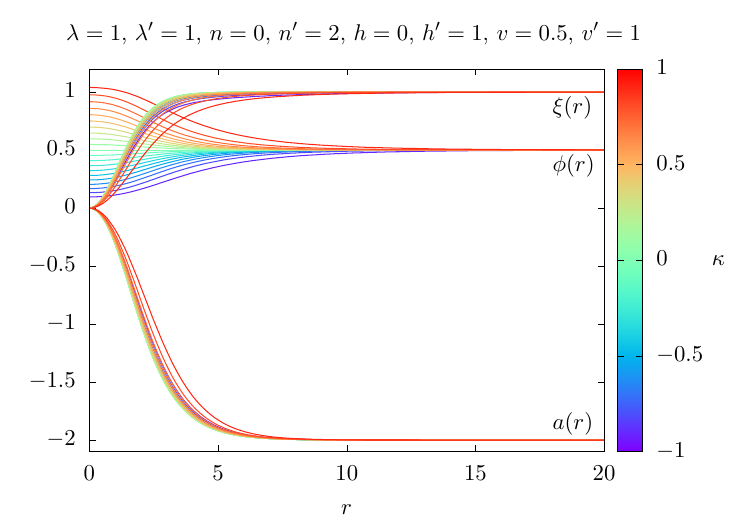}
\vspace*{-3mm}
\caption{\small String profiles for the fields $\phi(r)$, $\xi(r)$ and
$a(r)$ over the allowed range $| \kappa | < 1$. Due to $n=0$, $\phi(0)$
does not vanish, but there is still the condition $\phi'(0) = 0$.
\label{fign0np2}}
\end{center}
\vspace*{-6mm}
\end{figure}

We start with two cases where only one of the Higgs fields
performs a non-trivial winding, {\it i.e.}\ either $n=0$ or
$n'=0$. In these cases, we further set the gauge couplings
of the non-winding field to 0. There configurations
at $r=0$, $\xi(0)$ or $\phi(0)$ do not need to vanish
for $n'=0$ or $n=0$, respectively, in agreement with eq.\
(\ref{boundcore}). This is confirmed by the results shown in
Figures \ref{fign1np0} and \ref{fign0np2}, with $v=0.5$.

\begin{figure}[!h]
\vspace*{-2mm}
\begin{center}
\includegraphics[scale=0.8]{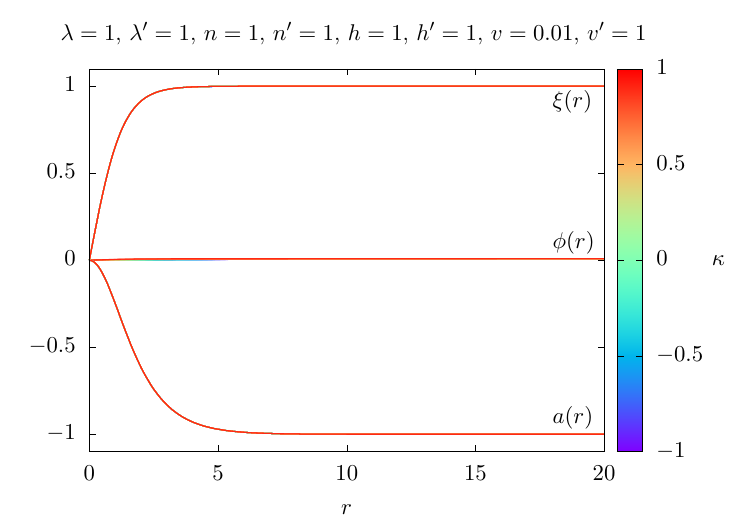}
\vspace*{-3mm}
\caption{\small String profiles for the fields $\phi(r)$, $\xi(r)$ and
$a(r)$ over the range $| \kappa | < 1$. We see a smooth and monotonous
behavior, similar to the prototype in Figure \ref{prot2prof}.\label{figsmooth}}
\end{center}
\vspace*{-6mm}
\end{figure}

We proceed to a simple example with windings in both Higgs fields,
$n = n' = 1$, where now the hierarchy condition is implemented as
$v' = 1 \gg v = 0.01$. As in the previous examples of Figures
\ref{fign1np0} and \ref{fign0np2},
Figure \ref{figsmooth} shows smooth and monotonous
transitions from $\phi(0) = \xi(0) = a(0) = 0$ to the asymptotic
values given in eq.\ (\ref{asymcond}). These curves hardly depend
on $\kappa$, because the potential is strongly dominated by the
purely $\chi$-dependent terms (note that
$\mu{'}^{2} = -\lambda' v{'}^{2} - \kappa v^{2}
\approx -\lambda' v{'}^{2}$ in this case).

\begin{figure}[!h]
\vspace*{-2mm}
\begin{center}
\includegraphics[scale=0.8]{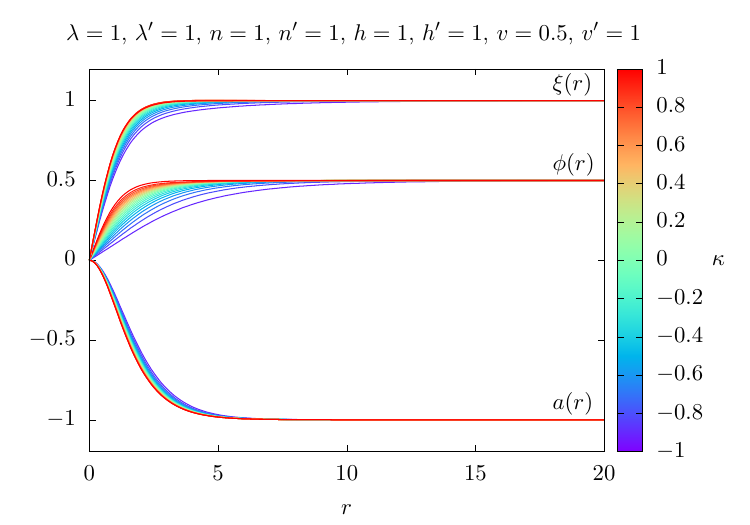}
\vspace*{-3mm}
\caption{\small String profiles for the fields $\phi(r)$, $\xi(r)$ and
$a(r)$ over the range $|\kappa | < 1$. The curves are still smooth
and monotonous, but we now see a significant dependence on $\kappa$.
\label{figkappa}}
\end{center}
\vspace*{-6mm}
\end{figure}

In order to further explore the scenario where the hierarchy
$v < v'$ is not that extreme, we return to the standard Higgs
VEV $v=1/2$, as in Figures \ref{fign1np0} and \ref{fign0np2},
while keeping all other parameters unchanged. The outcome is shown
in Figure \ref{figkappa}.
In this case, also the mixed term becomes relevant, as we see
from the significant $\kappa$-dependence at short distances.

\begin{figure}[!h]
\vspace*{-2mm}
\begin{center}
\includegraphics[scale=0.8]{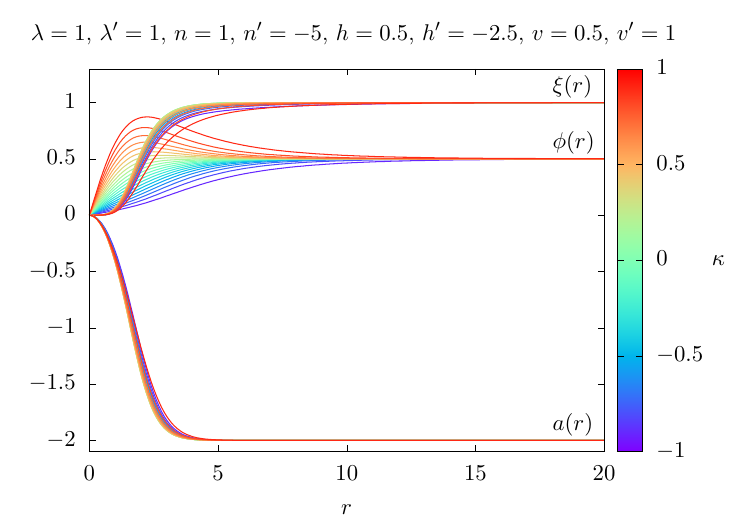}
\vspace*{-3mm}
\caption{\small String profiles for the fields $\phi(r)$, $\xi(r)$ and $a(r)$
  over the range $|\kappa | < 1$. For $\kappa \gtrsim 0$, we observe that the
  profile function $\phi (r)$ {\em overshoots} its vacuum expectation
  value in the vicinity of the string core.\label{fignpm5}}
\end{center}
\vspace*{-6mm}
\end{figure}

We proceed to the exploration of somewhat exotic scenarios by considering
a strong winding in the $\xi$-field, $n'=-5$. The other parameters and the
profiles are given in Figure \ref{fignpm5}.
We see that this leads to a qualitatively new feature: for increasing
radius $r$, the profile function $\phi(r)$ {\em overshoots} its asymptotic
value in cases where $\kappa$ is sufficiently positive. This
non-monotonous shape between the known asymptotic values at small
and large $r$ is amazing. An elementary particle, which
picks up some mass through the standard Higgs mechanism, and which
crosses such a cosmic string very close to its core, would temporarily 
increase its mass. However, even if such cosmic strings exist in the
Universe, this event does not seems obvious:
with the scale set according to eq.\ (\ref{units}), that particle would
have to cross the string at a distance $\lesssim 4 \cdot 10^{-3}\,
{\rm fm}$ from its core.

\begin{figure}[!h]
\vspace*{-2mm}
\begin{center}
\includegraphics[scale=0.8]{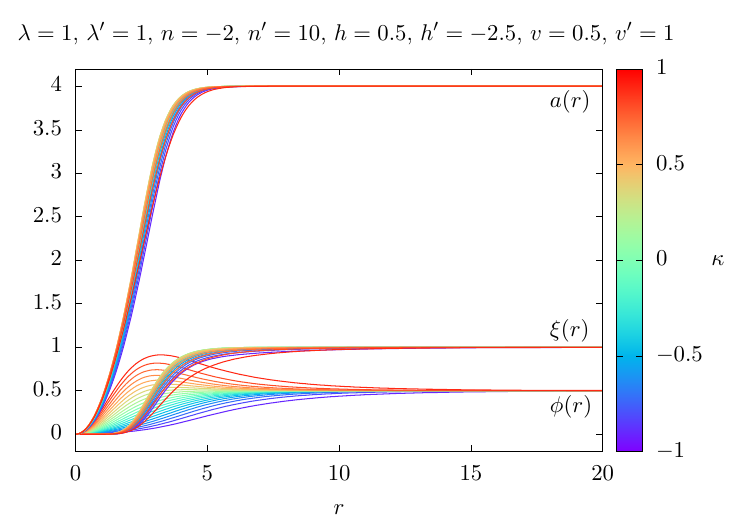}
\vspace*{-3mm}
\caption{\small String profiles for the fields $\phi(r)$, $\xi(r)$ and $a(r)$
over the range $|\kappa | < 1$. As in Figure \ref{fignpm5}, for
$\kappa \gtrsim 0$, we observe that the profile function $\phi (r)$
overshoots its vacuum expectation value in the vicinity of the string core.
\label{fignp10}}
\end{center}
\vspace*{-6mm}
\end{figure}

We now invert the windings and double their strengths, which yields the
profiles in Figure \ref{fignp10}. This drives $a(r \gg 1)$ up to 4,
while the overshooting behavior of $\phi(r)$ at $\kappa \gtrsim 0$
persists.

\begin{figure}[!h]
\vspace*{-2mm}
\begin{center}
\includegraphics[scale=0.8]{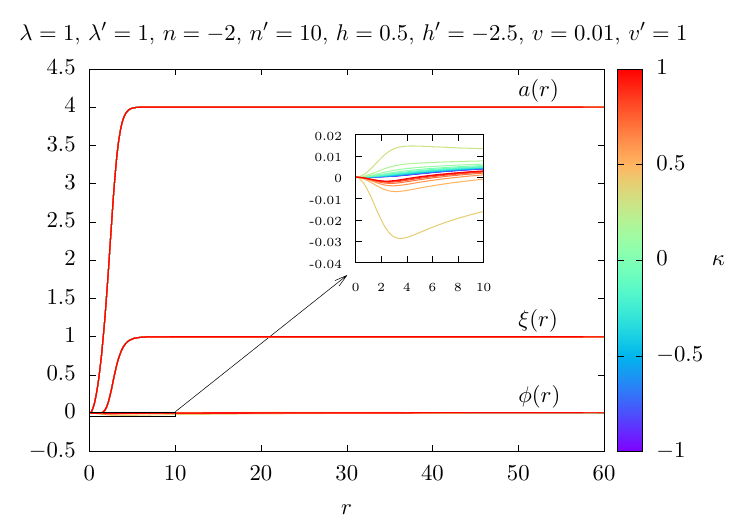}
\vspace*{-3mm}
\caption{\small For this parameter set, with the high winding $n'=10$,
we obtain a {\em co-axial} profile $\phi(r)$. For $\kappa < 0.25$,
we observe the typical behavior where $\phi(r)$ monotonously moves
to $v$. By gradually increasing $\kappa$ this profile first
overshoots, up to a discontinuity in $\kappa$ where $\phi(r)$
flips to a co-axial shape, which means that it takes a sign opposite
to $v$ near the core of the string.\label{figcoaxial}}
\end{center}
\vspace*{-6mm}
\end{figure}

Further numerical experiments reveal yet another surprising
phenomenon, which is even more exotic. An example is shown in
Figure \ref{figcoaxial}, with windings $n=-2$, $n'=10$, and
we restore the strong Higgs VEV hierarchy by setting $v=0.01$.
In this case, the profile function $\phi(r)$ is not only
non-monotonous, but for certain $\kappa$ values it is negative
at small $r$, {\it i.e.}\ it actually changes sign near the core.
We denote this kind of profile as {\em co-axial}.
It is hardly known that this remarkable behavior is possible,
although it was observed before by Bogomol'nyi, in the framework
of a standard type of cosmic strings \cite{Bogo}. In his study,
he showed that such strings can be unstable; here we leave the
question of stability for future investigations. Stability can be
explored numerically by means of variational techniques, along
the lines of Refs.\ \cite{Bogo,BogVain,JacReb,stabi}.

\begin{figure}
\vspace*{-2mm}
\begin{center}
\includegraphics[scale=0.8]{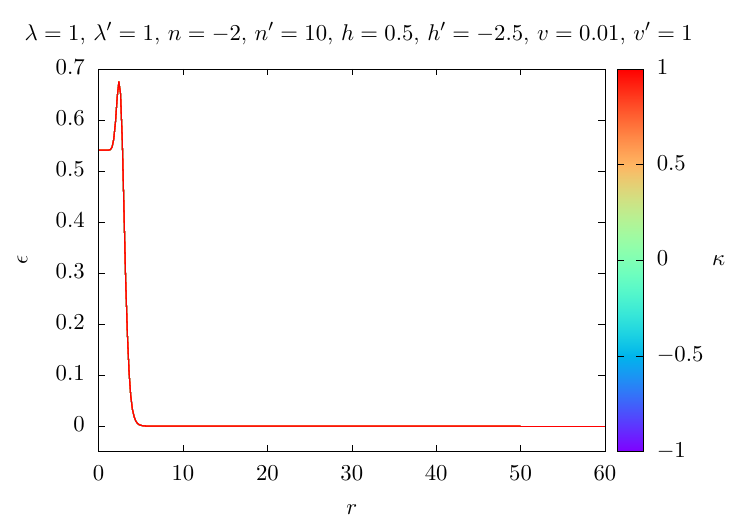}
\vspace*{-3mm}
\caption{\small Radial energy density $\eps$ of the solution for
the parameters of Figure \ref{figcoaxial}. It is obtained by a
straightforward generalization of eq.\ (\ref{proto2eps}).
We see a sharp peak in the co-axial regime near the core. In physical
units, $\eps =1$ corresponds to $4.79\cdot 10^{19} \, {\rm GeV/fm}^3$.
\label{edens}}
\end{center}
\vspace*{-6mm}
\end{figure}

For that case, Figure \ref{edens} shows the radial energy density
$\epsilon (r)$, which is computed by generalizing eq.\
(\ref{proto2eps}). In contrast to
the prototypes in Section \ref{string}, it does not monotonously
decrease towards 0, but in the co-axial domain it first shoots up
before performing its asymptotic decay. 

Let us now address the cases where our model can be embedded in an
SO(10) Grand Unified Theory \cite{Mink74}, as we announced in Section
\ref{model}. Relation (\ref{YpSO10}), along with the asymptotic
conditions for $a(r)$ given in eq.\ (\ref{asymcond}), requires
\be
\frac{h'}{h} = -5 \ , \qquad \frac{n'}{n} = -5 \ .
\ee
In fact, this is the case in the parameter sets of Figures
\ref{fignpm5}, \ref{fignp10}, \ref{figcoaxial} and \ref{edens}.

\begin{figure}
\vspace*{-2mm}
\begin{center}
\includegraphics[scale=0.8]{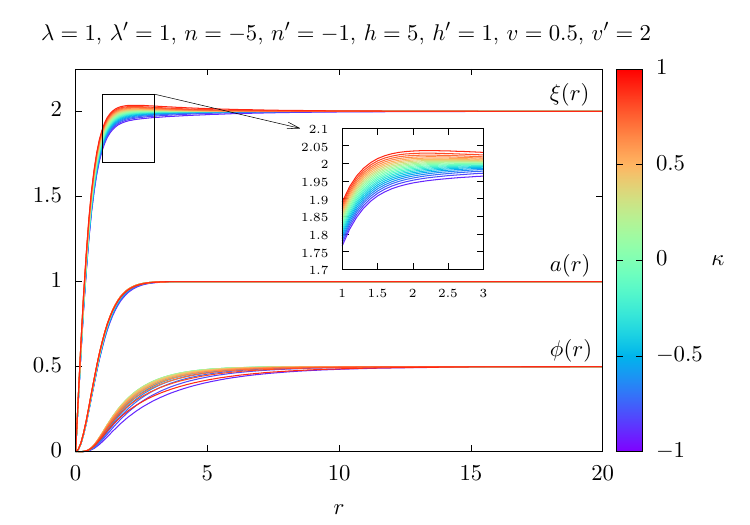}
\vspace*{-3mm}
\caption{\small Profile functions for a parameter set which leads to
the usual behavior for the profiles $\phi(r)$ and $a(r)$, but a slightly
overshooting behavior in this case of the profile function $\xi (r)$, around
$r \approx 1.5$, for $\kappa \gtrsim 0.6$.\label{figxiov}}
\end{center}
\vspace*{-6mm}
\end{figure}

At last, we show that also the profile $\xi(r)$ can overshoot at
short distances, as the example in Figure \ref{figxiov} shows.

\section{Overview and conclusions}

\label{conclu}
We studied a modest and well motivated extension of the Standard Model,
where the difference $B-L$ (baryon number minus lepton number)
is turned into the charge of a non-standard U(1) gauge symmetry.
In order to cancel gauge anomalies, we added right-handed neutrinos.
They obtain a mass by means of an additional 1-component Higgs field.

In the gauge-Higgs sector of this model, which includes the
standard and non-standard Higgs
field and the non-standard Abelian gauge field, we studied the
possible formation of cosmic strings for a variety of parameters
of the Lagrangian and winding numbers of both Higgs fields.
Low winding numbers tend to imply smooth and monotonous transitions
between the values of the (radial) profile functions in the core
and asymptotically far from it. For multiple windings, we observed
interesting and amazing effects, in particular the overshooting
of the profile function next to the core, and even a co-axial
behavior. In some cases, this formulation can be embedded into
an SO(10) GUT. Dark cosmic strings in a
similar set-up --- though with the inclusion of a neutral weak
gauge boson field --- were studied in Ref.\ \cite{HLV14}. That
work, however, does not report overshooting or co-axial (or other
non-monotonous) solutions for the profile functions.

One conceivable extension of this work could refer
to possible knot soliton solutions, along the lines of Ref.\
\cite{EHN24}, which involved a Chern-Simons term.

The question of the stability of our solutions to the field equations
is another aspect that we leave for future investigations,
based on a numerical variation study,
inspired by Refs.\ \cite{Bogo,BogVain,JacReb,stabi}.
In particular, Ref.\ \cite{Bogo,BogVain} observed
that Nielsen-Olesen type strings (cf.\ footnote \ref{fnAH}) with
$2 \lambda_{0}/g^{2} > 1$ are unstable for winding numbers $|n| > 1$.
On the other hand, they reported that for $2 \lambda_{0}/g^{2} < 1$
strings with any winding number could be stable, a property
which was substantiated in Ref.\ \cite{JacReb}.

For the parameter sets that we studied, the characteristic
radius of these strings (the range where the profile function
significantly deviates from the asymptotic behavior far from the
core) is several orders of magnitude below a nucleon radius.
They attain at  most ${\cal O}(10^{-3}) \, {\rm fm}$;
therefore they are good approximations to 1d ``Nambu-Goto strings''.

Regarding the {\em string tension}, we refer to the example of
the co-axial string profile illustrated in Figure \ref{figcoaxial},
and Figure \ref{edens} shows the corresponding energy density $\eps$.
The string tension is obtained by (numerically) integrating $\eps$ over
the plane perpendicular to the string (at fixed $z$-coordinate).
This yields the energy per unit length, {\it i.e.}\ the string tension,
$\mu \simeq 1.2 \times 10^{10} ~ {\rm GeV}^{2}$. It is usual to
express this quantity in a dimensionless form, by multiplying
with the gravitational constant $G$,
\be  \label{Gmubound}
G \mu \simeq 8.1 \times 10^{-29} \ .
\ee
Even if such a string has a length that corresponds to the diameter
of the visible Universe, about $28.5 ~{\rm Gpc}$, its mass is just
$9.5 \times 10^{25}~{\rm kg}$, slightly less than the mass of Neptune.

The methods how to derive observational bounds for cosmic are reviewed
{\it e.g.}\ in Refs.\ \cite{refbounds}. In particular, such bounds were
obtained from precise data of the Cosmic Microwave Background (CMB).
Cosmic strings may have caused 1d discontinuities
in the CBM (Kaiser-Stebbins-Gott effect). The direct search for such
1d discontinuities led to a bound around $G \mu < 4 \times 10^{-6}$
\cite{JS}. Statistical methods deal with the angular power spectrum
of CMB anisotropies. The analysis of data by the satellites
WMAP (Wilkinson Microwave Anisotropy Probe) and SDSS (Sloan Digital
Sky Survey) led to a bound of $G \mu < {\cal O} (10^{-7})$
\cite{WPW}. This magnitude was confirmed by an analysis of
Planck2015 data \cite{CACM}.

In recent years, this bound was significantly tightened based on gravitational
wave observations, referring specifically to (possible)
oscillating loops of cosmic strings:
Ref.\ \cite{BOS} reports $G \mu < 1.5 \times 10^{-11}$,
and Ref.\ \cite{LIGO3} even finds --- based on the third LIGO-Virgo-KARGA
Observing Run --- $G \mu < 4 \times 10^{-15}$. As an outlook, Ref.\ \cite{LISA}
predicts that the space-based observatory LISA (Laser Interferometer
Space Antenna) will arrive at a sensitivity to cosmic string networks
with string tensions $G \mu < {\cal O} (10^{-17})$.
However, even that bound is far from excluding the scenario of this work,
as our estimate (\ref{Gmubound}) shows. The discrepancy is so strong
that we can safely conclude that the other parameter sets considered
in this work are not observationally ruled out either, and won't be
excluded in the foreseeable future.

\ \\

\noindent
{\bf Acknowledgments:}
We thank Jo\~{a}o Pinto Barros and Uwe-Jens Wiese for inspiring
discussions and contributions to this project at an early stage.
We also thank Yu Hamada for valuable remarks about
related literature.
This work was supported by UNAM-DGAPA through PAPIIT projects IG100219
and IG100322, and by the Mexican {\it Consejo Nacional de Humanidades,
Ciencias y Tecnolog\'{\i}as} (CONAHCYT).

\appendix

\section{Numerical methods}

\label{numeri}
Our standard tool to obtain the solutions presented in Chapter
\ref{profile} was the Python function
{\tt scipy.integrate.solve$\_$bvp}. It applies the damped
Newton method to the functions of the radial coordinate $r$, which
is discretized in equidistant sites $\vec r = (r_0, r_1, \dots ,r_N)$,
where we typically used values of $N$ of order ${\cal O}(1000)$.
We inserted standard discretizations of the first and second derivatives.

The iteration of the straight Newton (or Newton-Raphson) method
may fail to converge to the correct solution ---
or fail to converge at all --- if the initial guess
is not sufficiently good, which represents a serious problem.
Effects like overshooting and oscillations in the iteration process
can be reduced by using the {\em damped} Newton method, which optimizes
the modification in the iteration step. This can be done by
minimizing the norm of the deviation from an exact solution.
Typically, this reduces the modification, which corresponds to
the damping effect.

For the initial guess, it turned out to be successful to insert an
exponential ansatz, {\it e.g.}\ for the standard Higgs profile function
it would take the form
\begin{equation} \label{profileansatz}
\phi (r) = (1 - e^{-r/r_0}) v \ , 
\end{equation}
where the scale $r_0$ can be varied (one might start from $r_0=1$).

For the generation of a set of solutions with varying parameter
$\kappa$, the solution for one $\kappa$-value is a good ansatz
for the next one, which differs only slightly. This procedure
also yields the overshooting and co-axial solutions, although the
original profile ansatz (\ref{profileansatz}) is monotonous.

The solutions were submit to a consistency test by inserting them
back into the discretized differential equations. The norm of the
deviation from an exact solution was at most of order ${\cal O}(10^{-3})$.
Such values only occurred at small $r$, where the system
is most delicate. In the range $r \geq {\cal O}(1)$, the precision
improved by further orders of magnitude.

As an alternative, we also implemented a relaxation method, which
iteratively applies a local improvement to the approximate solutions;
this usually amounts to a local smoothing.
Its convergence can be slow, but after a large number of iterations
(up to $10^7$), the results agree with the outcome of the damped
Newton method (starting from the same initial guess) within the
precision of the consistency test.

Finally, we also run experiments with the 4-point Runge-Kutta method,
although it is less appropriate for such boundary value problems.
For technical reasons (removable singularities), we cannot start at
$r=0$, so we started from $r_{\varepsilon}=0.01$, set the profile functions
at $r_{\varepsilon}$ to their theoretical values at $r=0$ and treated the
first derivatives at $r_{\varepsilon}$ as free parameters. Only after
fine-tuning them to a precision of $10^{-7}$ we obtained solutions, which
are asymptotically constant as $r$ becomes large within the range of
our plots in Chapter \ref{profile}. Then they agree again with the
solutions obtained by the damped Newton method and the relaxation method.
Varying the derivatives at $r_{\varepsilon}$, however, leads to a variety
of functions which diverge or oscillate as $r$ increases.

\section{Bound on the mixed term in the Higgs potential}

\label{bound}
We introduce the notation
\be
X = \Phi^{\dagger} \Phi \ , \quad Y = \chi^{*} \chi 
\ee
to rewrite the potential in the second line of eq.\ (\ref{PotPhiChi}) as
\be
V(X,Y) = \frac{\mu^{2}}{2} X + \frac{\mu'^{\, 2}}{2} Y
+ \frac{\lambda}{4} X^{2} + \frac{\lambda'}{4} Y^{2}
+ \frac{\kappa}{2} XY \ , 
\ee
with $\mu^{2}, \mu'^{\, 2} < 0$;\, $\lambda, \lambda'>0$ and $X,Y \geq 0$.

Clearly $V(0,0)=0$, $^{\lim}_{X \to + \infty} V = \, ^{\lim}_{Y \to + \infty} V =
+ \infty$. The Hessian matrix
$\left( \begin{array}{cc} \lambda/2 & \kappa/2 \\
\kappa /2 & \lambda'/2 \end{array} \right)$ is constant,
and it has the eigenvalues
\be
\varepsilon_{\pm} = \frac{1}{2} \left[ \frac{1}{2} (\lambda + \lambda ')
\pm \sqrt{\frac{1}{4} (\lambda + \lambda ')^{2} + \kappa^2 -
\lambda \lambda '} \, \right] \ .
\ee
The discriminant can be written as $(\lambda - \lambda')^{2}/4 + \kappa^{2}$,
which shows that imaginary eigenvalues are excluded. The above form
illustrates that $\varepsilon_{\pm} > 0$ holds if
\be  \label{kaplam}
\kappa^{2} < \lambda \lambda ' \ ,
\ee
which coincides with the constraint (\ref{kappa2constr}).
In this case, the function is convex everywhere, and thus bounded
from below.

There is a unique critical point in the $(X,Y)$-plane at
\be
(X_0,Y_0) = \frac{1}{\lambda \lambda' - \kappa^{2}}
  (\kappa \mu'^{\, 2} - \lambda' \mu^2, \kappa \mu^{\, 2} -
  \lambda \mu'^{\, 2}) \ .
\ee
Due to $\mu^{2}, \mu'^{\, 2} < 0$, this point is located in the
quadrant $X,\, Y \geq 0$ that we are considering (that can be
seen from the linear part of $V(X,Y)$).
This property is equivalent to condition (\ref{kappa1constr}).
With the constraint (\ref{kaplam}) it is a minimum,
which confirms that the potential is bounded from below.

It does not make sense to consider the case $\kappa^{2} = \lambda \lambda '$,
where $X_0$ and $Y_0$ diverge, and so do $v$ and $v'$, as eqs.\
(\ref{asymcond}) show.

For $\kappa^{2} > \lambda \lambda '$, we obtain $\varepsilon_{-} < 0$,
so $(X_0, Y_0)$ is a saddle point and the potential is not bounded
from below. Therefore our numerical study is restricted to
$\kappa$-values which fulfill the constraint (\ref{kaplam}).


\begin{thebibliography}{99}

\bibitem{PesSch} M.E.\ Peskin and D.V.\ Schroeder,
An Introduction to Quantum Field Theory, Addison Wesley, 1997.
  
\bibitem{CPT} W.\ Pauli,
On the Conservation of the Lepton Charge,
\href{https://doi.org/10.1007/BF02827771}
{Nuovo Cimento 6 (1957) 204-215.} \\
G.\ L\"{u}ders, Proof of the TCP theorem,
\href{https://doi.org/10.1016/0003-4916(57)90032-5}
{Ann.\ Phys.\ (N.Y.) 2 (1957) 1-15.} \\
R.\ Jost, Eine Bemerkung zum CTP-Theorem,
Helv.\ Phys.\ Acta 30 (1957) 409-416.
  
\bibitem{LIGO1}
B.P.~Abbott {\it et al.}\ (LIGO and Virgo Collaborations),
Constraints on cosmic strings using data from the first Advanced LIGO
observing run,
\href{https://doi.org/10.1103/PhysRevD.97.102002}
{Phys.\ Rev.\ D 97 (2018) 102002}.

\bibitem{LIGO3} R.~Abbott {\it et al.}\ (LIGO Scientific Collaboration,
  Virgo Collaboration and KAGRA Collaboration),
Constraints on Cosmic Strings Using Data from the Third
Advanced LIGO-Virgo Observing Run,
\href{https://doi.org/10.1103/PhysRevLett.126.241102}
{Phys.\ Rev.\ Lett.\ 126 (2021) 241102}.

\bibitem{Kibble} T.W.B.~Kibble,
Topology of cosmic domains and strings,
\href{https://iopscience.iop.org/article/10.1088/0305-4470/9/8/029}
     {J.\ Phys.\ A: Math.\ Gen.\ 9 (1976) 1387-1398};
Some implications of a cosmological phase transition,
\href{https://doi.org/10.1016/0370-1573(80)90091-5}
{Phys.\ Rep.\ 67 (1980) 183-199}.

\bibitem{HindKibVil} M.B.~Hindmarsh and T.W.B.~Kibble,
Cosmic Strings,
\href{https://iopscience.iop.org/article/10.1088/0034-4885/58/5/001}
{Rep.\ Prog.\ Phys.\ 58 (1995) 477-562}.\\
A.~Vilenkin and E.P.S.~Shellard,
Cosmic Strings and Other Topological Defects,
Cambridge University Press, 2000.

\bibitem{Bogo} E.B.\ Bogomol'nyi,
The stability of classical solutions,
Sov.\ J.\ Nucl.\ Phys.\ 24 (1976) 449-454.

\bibitem{Victor} V.~Mu\~{n}oz-Vitelly,
The structure of cosmic strings for a U(1) gauge field related to
the conservation of the baryon-number minus lepton-number,
M.Sc.\ thesis, Universidad Nacional Auton\'{o}ma de M\'{e}xico, 2022.

\bibitem{JA} J.A.~Garc\'{\i}a-Hern\'{a}ndez,
The profile of non-standard cosmic strings,
M.Sc.\ thesis, Universidad Nacional Auton\'{o}ma de M\'{e}xico, 2023.

\bibitem{proc} V.\ Mu\~{n}oz-Vitelly, J.A.\ Garc\'{\i}a-Hern\'{a}ndez
  and W.\ Bietenholz,
  The structure of cosmic strings of a U(1) gauge field for the
  conservation of $B - L$, 
\href{https://doi.org/10.31349/SuplRevMexFis.3.020713}
     {Supl.\ Rev.\ Mex.\ F\'{\i}s.\ 3 (2022) 020713}.

     
\bibitem{Abrikosov} A.A.\ Abrikosov,
 On the Magnetic Properties of Superconductors of the Second Group,
\href{http://www.w2agz.com/Library/Classic%20Papers%20in%20Superconductivity/Abrikosov,%20Type%20II%20Superconductivity,%20Sov-Phys%20JETP%205,%201174%20(1957).pdf} {Soviet Physics JETP 5 (1957) 1174};
The magnetic properties of superconducting alloys,
\href{https://doi.org/10.1016/0022-3697(57)90083-5}
{J.\ Phys.\ Chem.\ Sol.\ 2 (1957) 199}.
 
\bibitem{Ducci} S.\ Ducci, P.L.\ Ramazza, W.\ Gonz\'{a}lez-Vi\~{n}as
and F.T.\ Arecchi,
Order parameter fragmentation after a symmetry-breaking transition,
\href{https://doi.org/10.1103/PhysRevLett.83.5210}
{Phys.\ Rev.\ Lett.\ 83 (1999) 5210}.

\bibitem{Feyn} R.P.\ Feynman,
Application of Quantum Mechanics to Liquid Helium, {\it in}
\href{https://doi.org/10.1016/S0079-6417(08)60077-3}
{``Progress in Low Temperature Physics'',
ed.\ D.F.\ Brewer, North Holland, Amsterdam 1 (1955) 17-53}.

\bibitem{Lin} S.-Z. Lin {\it et al.},
Topological defects as relics of emergent continuous symmetry
and Higgs condensation of disorder in ferroelectrics,
\href{https://doi.org/10.48550/arXiv.1506.05021}
{Nat.\ Phys.\ 10 (2014) 970}.  

\bibitem{WBUrs} W.\ Bietenholz and U.\ Gerber,
Berezinskii-Kosterlitz-Thouless Transition and the Haldane Conjecture:
Highlights of the Physics Nobel Prize 2016,
\href{https://arxiv.org/abs/1612.06132}
{Rev.\ Cub.\ F\'{\i}s.\ 33 (2016) 156-168}.

\bibitem{Brand} R.H.~Brandenberger, A.C.~Davis and M.~Trodden,
Cosmic strings and electroweak baryogenesis,
\href{https://doi.org/10.1016/0370-2693(94)91402-8}
{Phys.\ Lett.\ B 335 (1994) 123-130}.

\bibitem{James} M.~James, L.~Perivolaropoulos and T.~Vachaspati,
Detailed stability analysis of electroweak strings,
\href{https://doi.org/10.1016/0550-3213(93)90046-R}
{Nucl.\ Phys.\ B 395 (1993) 534}.\\
H.~Weigel, M.~Quandt and N.~Graham, Stable charged cosmic strings,
\href{https://doi.org/10.1103/PhysRevLett.106.101601}
{Phys.\ Rev.\ Lett.\ 106 (2011) 101601}.

\bibitem{Zurek} W.H.\ Zurek,
Cosmological experiments in superfluid helium?,
\href{https://doi.org/10.1038/317505a0}{Nature \textbf{317} (1985) 505}.

\bibitem{JS} E.\ Jeong and G.F.\ Smoot, Search for Cosmic Strings
in Cosmic Microwave Background Anisotropies,
\href{https://doi.org/10.1086/428921}
{Astrophys.\ J.\ 624 (2005) 21}.

\bibitem{WPW} M.\ Wyman, L.\ Pogosian and I.\ Wasserman,
Bounds on cosmic strings from WMAP and SDSS,
\href{https://doi.org/10.1103/PhysRevD.72.023513}
{Phys.\ Rev.\ D72 (2005) 023513}
[Erratum: \href{https://doi.org/10.1103/PhysRevD.73.089905}
  {Phys.\ Rev.\ D73 (2006) 89905(E)}].

\bibitem{CACM} T.~Charnock, A.~Avgoustidis, E.J.~Copeland
  and A.~Moss,
CMB constraints on cosmic strings and superstrings,
\href{https://doi.org/10.1103/PhysRevD.93.123503}
{Phys.\ Rev.\ D 93 (2016) 123503}.

\bibitem{JelCug} A.\ Jeli\'{c} and L.F.\ Cugliandolo,
Quench dynamics of the 2d XY model,
\href{https://doi.org/10.1088/1742-5468/2011/02/P02032}
{J.\ Stat.\ Mech.\ Theory Exp.\ 2011 (2011) P02032}.

\bibitem{Lat23} E.\ L\'{o}pez-Contreras, J.F.\ Nieto Castellanos,
E.N.\ Polanco-Eu\'{a}n and W.\ Bietenholz,
Non-equilibrium dynamics of topological defects in the 3d O(2) model,
\href{https://doi.org/10.22323/1.453.0346}
{PoS LATTICE2023 (2024) 346}.\\
J.F.\ Nieto Castellanos,     
Topological defects in the O(2) model out of equilibrium,
M.Sc.\ thesis, Universidad Nacional Auton\'{o}ma de M\'{e}xico, 2024.\\
E.\ L\'{o}pez-Contreras,
Topogical defects and the Kibble-Zurek mechanism in the 3d O(2) model,
M.Sc.\ thesis, Universidad Nacional Auton\'{o}ma de M\'{e}xico,
2025.


\bibitem{ISY} M.~Ibe, M.~Suzuki and T.T.~Yanagida,
$B-L$ as a Gauged Peccei-Quinn Symmetry,
\href{https://doi.org/10.1007/JHEP08(2018)049}
{JHEP 08 (2018) 049}.

\bibitem{Weinberg} S.~Weinberg,
The Quantum Theory Of Fields. Vol. 2: Modern
Applications, Cambridge University Press, 1996. 

\bibitem{ADH03} T.~Appelquist, B.A.~Dobrescu and A.R.~Hopper,
Nonexotic neutral gauge bosons,
\href{https://doi.org/10.1103/PhysRevD.68.035012}
{Phys.\ Rev.\ D 68 (2003) 035012}.

\bibitem{Mink74} H.~Fritzsch and P.~Minkowski,
Unified interactions of leptons and hadrons,
\href{https://doi.org/10.1016/0003-4916(75)90211-0}
{Ann.\ Phys.\ (NY) 93 (1975) 193-266}.

\bibitem{Buch91} W.~Buchm\"{u}ller, C.~Greub and P.~Minkowski,
Neutrino masses, neutral vector bosons and the scale of $B-L$ breaking,
\href{https://doi.org/10.1016/0370-2693(91)90952-M}
{Phys.\ Lett.\ B 267 (1991) 395-399}.


\bibitem{BogVain} E.B.\ Bogomol'nyi and A.I.\ Vainshtein,
Stability of strings in gauge Abelian theory,
Sov.\ J.\ Nucl.\ Phys.\ 23 (1976) 588-591.

\bibitem{JacReb} L.\ Jacobs and C.\ Rebbi,
Interaction energy of superconducting vortices,
\href{https://doi.org/10.1103/PhysRevB.19.4486}
     {Phys.\ Rev.\ B 19 (1979) 4486}.

\bibitem{stabi} T.\ Vachaspati and A.\ Ach\'{u}carro,
Semilocal cosmic strings,
\href{https://doi.org/10.1103/PhysRevD.44.3067}
{Phys.\ Rev.\ D 44 (1991) 3067}.\\
M.\ Hindmarsh,
Existence and stability of semilocal strings,
\href{https://doi.org/10.1103/PhysRevLett.68.1263}
     {Phys.\ Rev.\ Lett.\ 68 (1992) 1263}.
  

\bibitem{HLV14} J.M.~Hyde, A.J.~Long and T.~Vachaspati,
Dark strings and their couplings to the standard model,
\href{https://doi.org/10.1103/PhysRevD.89.065031}
{Phys.\ Rev.\ D 89 (2014) 065031}.
  
\bibitem{EHN24} M.~Eto, Y.~Hamada and M.~Nitta,
Tying knots in particle physics,
\href{https://arxiv.org/pdf/2407.11731}
{arXiv:2407.11731 [hep-ph]]}.

\bibitem{refbounds} T.\ Vachaspati, L.\ Pogosian and D.\ Steer,
Cosmic strings,
\href{https://arxiv.org/pdf/1506.04039}
{Scholarpedia 10 (2015) 31682}.\\
L.\ Sousa,
Cosmic strings and gravitational waves,
\href{https://doi.org/10.1007/s10714-024-03293-x}
{Gen.\ Relativ.\ Gravit.\ 56 (2024) 105.}

\bibitem{BOS} J.J.\ Blanco-Pillado, K.D.\ Olum and X.\ Siemens,
New limits on cosmic strings from gravitational wave observation,
\href{https://doi.org/10.1016/j.physletb.2018.01.050}
{Phys.\ Lett.\ B 778 (2018) 392-396.}

\bibitem{LISA}
P.\ Auclair {\it et al.},
Probing the gravitational wave background from cosmic strings
with LISA,
\href{https://doi.org/10.1088/1475-7516/2020/04/034}
{J.\ Cosmol.\ Astropart.\ Phys.\ 04 (2020) 034}.

\end{thebibliography}
\end{document}